\documentclass[prb,preprint,showpacs]{revtex4}

\usepackage{graphicx}
\usepackage{color}
\usepackage{amsmath}
\usepackage{amssymb}
\usepackage{rotating}
\usepackage{bm}

\begin{document}

\title{Elementary transitions and magnetic correlations in  
two-dimensional disordered nanoparticle ensembles} 

\author{G. M. Pastor}
\affiliation{
Institut f\"ur Theoretische Physik, Universit\"at Kassel, 
Heinrich Plett Stra\ss e 40, 34132 Kassel, Germany} 

\author{P. J. Jensen} 
\affiliation{Institut f\"ur Theoretische Physik, Freie Universit\"at 
Berlin, Arnimallee 14, 14195 Berlin, Germany}

\begin{abstract} 
The magnetic relaxation processes in disordered two-dimensional 
ensembles of dipole-coupled magnetic nanoparticles are theoretically
investigated by performing numerical simulations.
The energy landscape of the system is 
explored by determining saddle points, adjacent local minima, energy 
barriers, and the associated minimum energy paths (MEPs) as functions 
of the structural disorder and particle density. 
The changes in the magnetic order of the nanostructure along the 
MEPs connecting adjacent minima are analyzed from a local perspective.  
In particular, we determine the extension of the correlated region
where the directions of the particle magnetic moments vary 
significantly. It is shown that with increasing degree of disorder 
the magnetic correlation range decreases, i.e., 
the elementary relaxation processes become more localized. 
The distribution of the energy barriers, and their relation to the 
changes in the magnetic configurations are quantified. 
Finally, some implications for the long-time magnetic relaxation 
dynamics of nanostructures are discussed.
\end{abstract}


\maketitle

\section{Introduction}

Two-dimensional (2D) ensembles of interacting magnetic nanoparticles 
are currently the subject of an intense fundamental and technological 
research activity.\cite{Respaud98,Dorman97}  
The magnetic particles in these systems are subject to both local 
couplings and long-range interactions. 
By varying the interparticle distance the relative importance of 
these two contributions can be tuned to a certain extent. 
The magnetic response can then be correlated with the size distribution
and the geometrical arrangement of the nanoparticles, the latter
being experimentally accessible through a number of microscopy techniques.
Of particular interest is the slow magnetic dynamics,\cite{Lottis91} 
like magnetic ageing, rejuvenation, or magnetic viscosity.\cite{Jonsson95}
These non-equilibrium effects are reminiscent of the spin glass behavior 
observed in impurity spin systems with nonuniform exchange 
couplings.\cite{Binder86} In fact, the magnetic nanostructures
studied in this paper can be viewed as a super-spin glass with giant 
magnetic moments that interact via the dipole coupling. 

One of the experimental approaches to 2D nanostructures derives from 
investigations of three-dimensional (3D) ferrofluids consisting of 
non-overlapping magnetic particles in a liquid or frozen solvent.\cite{Luo91}
The controlled preparation of 2D nanoparticle ensembles with different 
degrees of structural and magnetic disorder has recently become 
feasible by reducing the thickness of 3D ferrofluids to within a few 
particle diameters.\cite{Respaud98,Kleemann01,Binns02}
A second experimental approach to 2D magnetic nanosystems stems from  
the study of thin magnetic films that are grown by depositing atoms 
on nonmagnetic substrates.\cite{Bland94,brune,wissen,bansmann} 
In the early stages of film 
growth, before a continuous layer is formed, it is possible to
obtain a controlled distribution of magnetic islands that 
interact by long-range forces like dipole 
coupling or Ruderman-Kittel-Kasuya-Yosida (RKKY) interactions 
in case of metallic substrates.\cite{Bruno95} 
As a consequence of the preparation conditions, these 
magnetic nanostructures are usually characterized by the presence of
a considerable degree of structural disorder. This manifests itself, 
for instance, in the particle-size dispersion 
or in the irregularity of the particle arrangement. 

The magnetic behavior of magnetic nanostructures is determined by 
the interplay between single-particle properties 
(e.g., magneto-crystalline or shape anisotropies) 
and interparticle effects (e.g., dipolar, RKKY or exchange interactions). 
For weakly coupled systems the local contributions dominate,
and the interactions can be handled as perturbations.
In this case the magnetic relaxation is governed by single-particle 
fluctuations that can be described by a
superparamagnetic Arrhenius-like model 
showing blocking effects at low temperatures.\cite{SaB00} 
However, in the most interesting case of strongly interacting magnetic
nanostructures, for example, for densely packed nanoparticles, 
the single-particle approach is no longer appropriate.\cite{Dorman97}
Here the system exhibits a complicated, spin-glass-like
behavior, characterized by a complex energy landscape, a large number
of metastable minima, and a non-collinear magnetization of the
particle ensemble.\cite{IgL03,JeP02} For these strongly interacting systems 
the magnetization dynamics is determined by a collective response. 
In other words, changes in the magnetization direction of 
a given particle inevitably cause neighboring particles to
vary their magnetic directions as well. During the 
transition from a local energy-minimum to another one, 
within a magnetic relaxation process,  
the magnetization directions of many particles change. 
Therefore, the collective behavior of the nanostructure
has to be taken into account from the start.

The disorder within the nanostructure and the  
anisotropic long-ranged nature of the dipole interaction 
represent a serious challenge for theoretical investigations so that
simple analytical approaches are in general not applicable.
Therefore, numerical simulations have been 
performed in order to achieve a detailed description of 
these systems.\cite{Kechrakos98} Phenomenological, macroscopic 
quantities like the field-cooled and zero-field-cooled 
magnetization curves, the 
linear and nonlinear susceptibilities, or the magnetic relaxation 
rate have been calculated.\cite{Binns02,Zaluska96,Chantrell00,Dorman88}
Moreover, the magnetic transitions between energy minima have been 
investigated by solving a continuum dynamical model 
including dissipation and thermal fluctuation terms 
(Langevin dynamics) as well as by using path-integral methods.\cite{Ber98}
In order to take into account the diversity of nonequivalent trajectories,
these calculations involve optimizations in a space of dimension
$d=n\nu k$, where $n$ is the number of particles, $\nu$ the number 
of degrees of freedom, and $k$ the number discretizations of the 
relaxation path.

The main purpose of this paper is to investigate the 
collective microscopic processes occurring along the 
magnetic relaxations of disordered magnetic nanostructures. 
To this aim we focus on the elementary transitions along the 
minimum energy path (MEP) connecting two neighboring local minima 
across a single saddle point. These fundamental processes 
can be regarded as the microscopic analogy to the Barkhausen-jumps 
of the magnetization in hysteresis loops. The actual macroscopic
relaxation is the result of a succession of such transitions 
between metastable states. Understanding them is therefore very important 
for the modelization of the long-time dynamics. One of the goals of 
this work is to analyze the properties of these magnetic 
rearrangements from a local perspective by determining the variations 
$\Delta\phi_i$ in the directions of the magnetic moments 
at each nanoparticle $i$. In particular a correlation 
range $\lambda_\mathrm{c}$ is defined, that measures the spatial extent 
of the relevant collective processes as a function of experimentally 
important parameters like particle density and degree of 
disorder. Furthermore, the distribution of the energy 
barriers is quantified in order to determine how they correlate with 
the changes in the magnetic configurations.  

The remainder of the paper is organized as follows. 
In Sec.~\ref{sec:mod} the theoretical background 
describing the interacting magnetic nanoparticle system is presented. 
The methods for calculating saddle points, adjacent minima, 
relaxation paths and the resulting changes in the magnetic 
order along elementary transitions are outlined. 
In Sec.~\ref{sec:res} representative results for the
elementary transitions and the associated correlation ranges 
are reported and discussed. The distribution of the 
energy barriers, and their relation to the variation in the 
magnetic configurations are investigated. Finally, we conclude  
in Sec.~\ref{sec:conc} by pointing out some of the potential 
implications of this study for magnetization reversal dynamics 
and high-density magnetic recording.

\section{Model and calculation method}
\label{sec:mod}

The purpose of this section is to describe the main aspects of the 
present model of disordered magnetic nanostructures. Further 
details of the theory may be found in Ref.~\onlinecite{JeP02}. We 
consider a 2D system of $N$ non-overlapping, spherical magnetic 
particles contained in a square unit cell with periodic boundary 
conditions. Due to the strong interatomic ferromagnetic exchange 
interactions and the small size of the particles under consideration, 
it is clear that each particle $i$ can be 
regarded as a single-domain or Stoner-Wohlfarth magnet
carrying a giant spin $\mathbf{m}_i$.\cite{StW48}
The modulus of the local moments is given by 
$m_i = n_i\,\mu_{at}$, where $\mu_{at}$ is the atomic magnetic moment 
and $n_i$ the number of atoms in particle $i$. Different kinds of 
structural arrangements are considered in the following: 
(i) a periodic square-lattice array, 
(ii) a disordered array where the particles are randomly
dispersed around the sites of a periodic square lattice according to 
a Gaussian distribution with standard deviation $\sigma_R$, and 
(iii) a fully-random particle distribution of non-overlapping particles 
within the unit cell. 

In this paper we focus on the effects of the dipole coupling on the
collective magnetic behavior. This is a particularly interesting
physical situation, since in the limit of weak magnetic anisotropy and 
external magnetic field the numbers of local minima and the complexity of
the energy landscape are expected to be largest. 
For each particle arrangement the dipolar interaction 
between the particle moments $\mathbf{m}_i$ is given by 
\begin{equation} \label{e1} 
E = \frac{\mu_0}{2}\;\sum_{i\ne j} 
\left[\mathbf{m}_i\cdot \mathbf{m}_j\;r_{ij}^{-3}-
3\,\big(\mathbf{r}_{ij}\cdot \mathbf{m}_i 
\big)\,\big(\mathbf{r}_{ij}\cdot \mathbf{m}_j\big)\; r_{ij}^{-5}\right] 
\;, \end{equation} 
where $\mathbf{r}_{ij} = \mathbf{r}_i - \mathbf{r}_j$ is
the vector between the centers of particles $i$ and $j$,  
$r_{ij}=|\mathbf{r}_{ij}|$ the corresponding distance, and 
$\mu_0$ the vacuum permeability. The infinite range of the dipole 
interaction is taken into account by computing an Ewald-type summation 
over all periodically arranged unit cells of the extended planar 
system.\cite{Jen97} Since the particles do not have direct metal-metal 
contacts, no interparticle exchange interactions are included. 
Taking into account these short-range contributions, as well as local 
magnetic anisotropies or external magnetic fields, poses no major technical 
difficulties and could be investigated by similar simulations. 
For simplicity we assume that all particles have the same size
$n_i=30000$ atoms, which corresponds to a particle radius of about 15 lattice 
constants. Previous studies have actually shown that the dominant disorder 
effects are due to the irregularity of the particle arrangement
and that the distribution of particle sizes yields a similar 
behavior.\cite{JeP02}
The directions of the particle magnetizations $\mathbf{m}_i$ 
are restricted to the plane of the nanostructure. Hence, the magnetic 
configurations are characterized for simplicity by the set of in-plane 
angles $\{\phi_i\}$ in the unit cell ($i=1,\ldots N$).

The particle ensemble is characterized by the number of particles $N$ 
in the unit cell, by the standard deviation $\sigma_R$ of the structural 
disorder of the particle array, and by the surface coverage $C$.\cite{JeP02} 
The size of the unit cell is defined in terms of $N$ and the average 
interparticle distance $R_0$. For a given realization of the nanostructure 
we sample randomly the energy landscape and determine a large number
of first-order saddle points (SPs) or transition states. These are
characterized by having one negative eigenvalue of the Hessian of 
the interaction energy $E$ [see Eq.~(\ref{e1})]. To locate the SPs 
we use a variant of the eigenvector-following 
method that is based on the eigenvalues and eigenvectors of the Hessian
matrix as described in the Appendix. The two neighboring energy minima 
adjacent to each SP and the associated MEP are 
obtained by a simple steepest descent procedure starting 
at the SP along the negative-curvature direction.
In this way the elementary relaxation path, the energies $E_\mathrm{S}$ 
and $E_\mathrm{M_{1,2}}$ for the saddle point $S$ and adjacent minima $M_1$ 
and $M_2$, as well as the corresponding sets of magnetization angles 
$\{\phi_i(S)\}$ and $\{\phi_i(M_{1,2})\}$, are determined. In this
context it is interesting to mention that loop-like paths also exist, 
where the SP connects a local minimum with itself via a closed loop
(i.e., $M_1=M_2$). The presence of these features of the energy landscape 
is favored by the periodicity of the angular functions entering 
Eq.~(\ref{e1}). Nevertheless, loop-like MEPs have been excluded from 
the present analysis, since they do not contribute to the magnetic relaxation. 

The range $\lambda_\mathrm{c}$ of the magnetic correlations involved 
in an elementary transition is quantified in two steps as follows. 
First, the epicenter $\mathbf{R}_\mathrm{c}$ of the angular variations 
$\Delta\phi_i = |\phi_i(\mathrm{M_1}) - \phi_i(\mathrm{M_2})|$
between the two adjacent minima is determined from 
\begin{equation} \label{e3}
\mathbf{R}_\mathrm{c}=\frac{\sum_i \Delta\phi_i
\;\mathbf{r}_i}{\sum_i \Delta\phi_i
} \;, 
\end{equation} 
where $\mathbf{r}_i$ is the position of the center of particle $i$. 
Note that $\mathbf{R}_\mathrm{c}$ needs not to coincide with the 
position of any particle. Second, the dependence of $\Delta\phi_i$ on 
the distance $|\mathbf{r}_i-\mathbf{R}_\mathrm{c}|$ to the epicenter
is fitted by an exponential of the form
\begin{equation} \label{el}
\Delta\phi_i = \Delta\phi_\mathrm{max} \; 
\exp(-2\,|\mathbf{r}_i-\mathbf{R}_\mathrm{c}|/\lambda_\mathrm{c}) \;,
\end{equation}
from which the correlation range $\lambda_\mathrm{c}$ is obtained. 
The angular variations are normalized with respect to the maximal 
change $\Delta\phi_\mathrm{max}$ occurring in that particular pair 
of states. The factor 2 in the exponential is introduced in order 
that $\lambda_\mathrm{c}$ covers most of the significant angular 
variations. Hence, for $|\mathbf{r}_i-\mathbf{R}_\mathrm{c}| = 
\lambda_\mathrm{c}$ the change of angle $\Delta\phi_i$ is reduced by 
a factor $e^2$ with respect to $\Delta\phi_\mathrm{max}$.
However, the use of an exponential fit should not be interpreted as a 
statement on the precise distance dependence of $\Delta\phi_i$, which 
is presently not known in detail. 

The Euclidean distance between the saddle point $S$ and the adjacent 
minimum $M$ is given by
\begin{equation} \label{e2} 
D_\mathrm{SM}=\{1-\frac{1}{N}\sum_{i=1}^N\;
\cos\big[\phi_i(\mathrm{S})-\phi_i(\mathrm{M})\big] \}^{1/2} \; .
\end{equation} 
$D_\mathrm{SM}$ provides a quantitative measure of the changes
in the magnetic order during an elementary transition. The 
correlation ranges $\lambda_\mathrm{c}$ and distances $D_\mathrm{SM}$ 
are averaged over a large number of elementary transitions 
(typically a few thousands) and several geometrical arrangements 
of the particles in the nanostructure. 
In the following calculations parameters appropriate 
for Fe particles are used, i.e., $\mu_{at}=2.2\;\mu_B$
and nearest neighbor distance $a_0=2.5$~\AA.

\section{Results and Discussion}
\label{sec:res}

The most stable magnetic arrangement of a periodic square lattice of 
dipole coupled magnetic nanoparticles having all the same size is know 
to correspond to the so-called microvortex state.\cite{JeP02} 
This magnetic order has a vanishing net magnetization 
and is characterized by a continuous degeneracy with respect to the 
microvortex angle. Any, however small, deviation from 
the square-lattice symmetry lifts this degeneracy. For
weak disorder the low-energy states preserve a close resemblance 
with a microvortex state but as the disorder increases
the apparent extended periodicity of the magnetic structure 
is completely lost. In the case of strong disorder
the magnetic state is characterized by 
head-to-tail arrangements of the particle moments and by 
local flux closures of the magnetization profile. In this Section 
results are presented and discussed for the correlation range 
$\lambda_\mathrm{c}$, the energy barriers 
$\Delta E_\mathrm{SM}=E_\mathrm{S}-E_\mathrm{M}$, and the 
distances $D_\mathrm{SM}$, including averages over many elementary 
transitions as well as different values of the parameters $N$, $\sigma_R$ 
and $C$ characterizing the magnetic nanostructure.

\begin{figure}
\includegraphics[width=9.5cm]{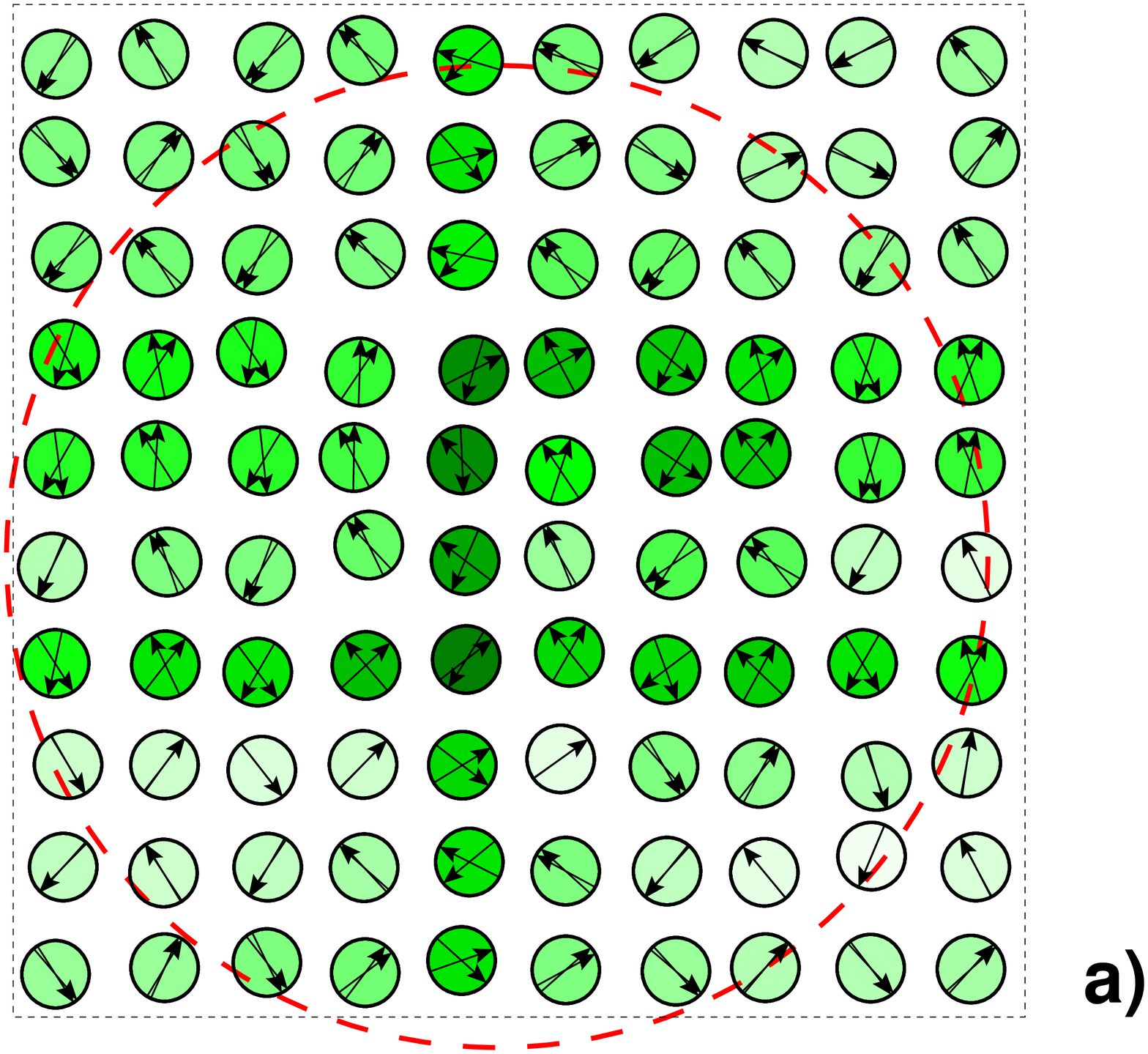} \\[0.3cm] 
\includegraphics[width=9.5cm]{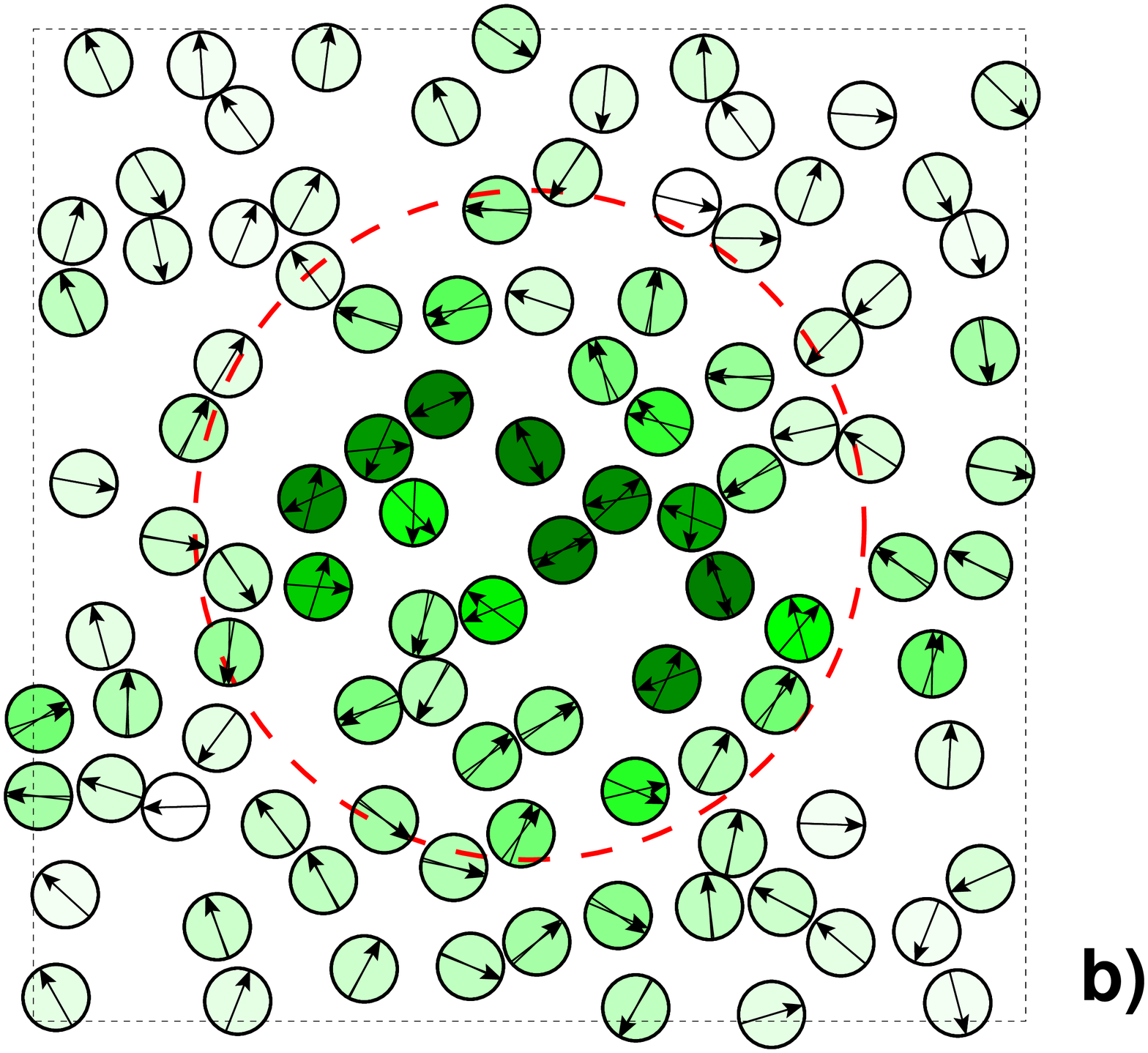} 
\caption{ \label{fig1} 
(Color online) Illustration of elementary transitions in 
dipole-coupled nanostructures having a) a weak disorder with 
$\sigma_R=5 \%$, and b) a random particle distribution. In both cases 
the unit cell (dotted frame) contains $N=100$ particles and the 
coverage is $C=0.35$. The arrows 
show the directions of the magnetic moments corresponding 
to two adjacent local minima that are separated by a single 
energy barrier. The shading indicates the importance of the changes 
in the direction of the local moments. The dashed circle refers to 
the correlation range $\lambda_\mathrm{c}$ where the angular 
variations are most significant. 
        } 
\end{figure} 

Figure~\ref{fig1} illustrates typical examples of elementary 
magnetic transitions. The arrows refer to the magnetic 
directions of two adjacent minima, and the different shadings 
indicate the importance of the associated angular changes
$\Delta\phi_i$. Two different structural arrangements are considered:
a) an almost periodic square-lattice array with $\sigma_R=5\%$, 
and b) a fully-random particle setup. The resulting correlation range 
$\lambda_\mathrm{c}$ is indicated by a dashed circle. One clearly 
observes that $\lambda_\mathrm{c}$ is significantly larger for the 
almost periodic arrangement than for the random ensemble. In fact, in 
the former case $\lambda_\mathrm{c}$ is often 
comparable with the size of 
the unit cell. Hence, to obtain reliable quantitative results one 
would have to consider a much larger cell so that spurious effects 
from the periodic boundary conditions are avoided. In contrast, for 
strong disorder the region with significant orientational changes of 
$\mathbf{m}_i$ remains quite small. In this case the
results for $\lambda_\mathrm{c}$ are not affected by considering 
even larger unit cells. 
The magnetic transition presented in Figure~\ref{fig1}(b) 
refers to a relatively large $\lambda_\mathrm{c}$, usually the 
regions of significant magnetic variations for this kind of 
arrangement are somewhat smaller. Note that for small disorder, 
Figure~\ref{fig1}(a), the magnetic variations are not isotropic but 
still reflect the square symmetry. 

\begin{figure}
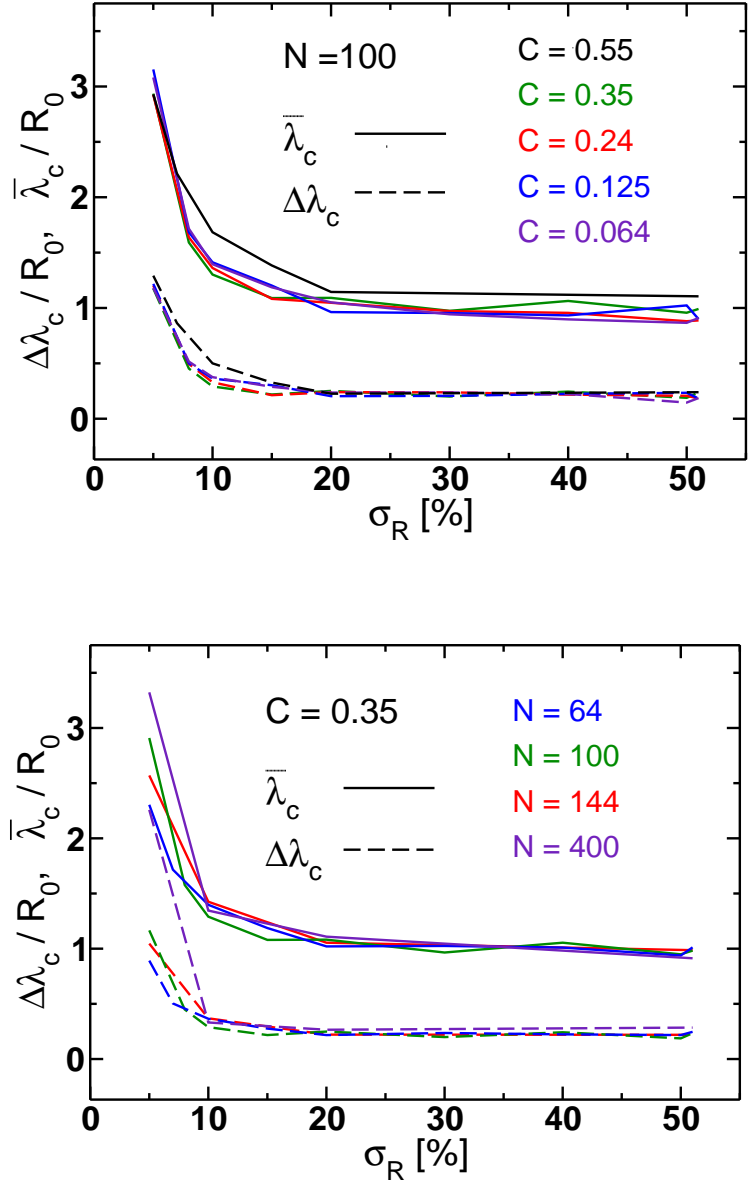

\begin{center}
\includegraphics[scale=0.413]{fig2a.eps} \\[13mm] 
\includegraphics[scale=0.4]  {fig2b.eps}
\vspace{-5mm}
\end{center} 
\caption{ \label{fig3} 
(Color online) Average correlation range 
$\overline{\lambda}_\mathrm{c}$ and standard deviation 
$\Delta\lambda_\mathrm{c}$ as functions of the structural disorder 
$\sigma_R$. The results are obtained by sampling a large number 
of elementary transitions. In the top figure different coverages 
$C$ are considered, while in the bottom figure the number of particles 
$N$ in the unit cell is varied. 
        } 
\end{figure} 

In order to derive a quantitative estimate of $\lambda_\mathrm{c}$ we 
approximate the dependence of $\Delta\phi_i$ on 
$|\mathbf{r}_i - \mathbf{R}_\mathrm{c}|$ by an exponential law 
[see Eq.~(\ref{el})]. Figure \ref{fig3} shows the correlation range 
$\overline{\lambda}_\mathrm{c}$  
averaged over a large number of elementary transitions 
as a function of the structural disorder $\sigma_R$. 
A single realization of the nanostructure is considered
either for different coverages $C$ keeping the number $N$ 
of particles in the unit cell constant, or vice versa. 
Results are given for $\overline{\lambda}_\mathrm{c}$ 
and for the standard deviation 
$\Delta\lambda_\mathrm{c} = \sqrt{\langle (\lambda_\mathrm{c} - 
\overline{\lambda}_\mathrm{c})^2 \rangle}$ in units of the average 
interparticle distance $R_0$. Notice that $R_0$ increases as the 
coverage $C$ is reduced. 
One observes that both $\overline{\lambda}_\mathrm{c}$ and
$\Delta\lambda_\mathrm{c}$ decrease rapidly with increasing 
$\sigma_R$. Already for $\sigma_R>10$--$20 \%$ their values  
are practically indistinguishable from those of a random particle 
ensemble. Usually $\overline{\lambda}_\mathrm{c} / R_0$ does not depend 
strongly on the coverage $C$, except for the largest values 
(e.g., $C=0.55$) where the non-overlap constraint and the assumed
small size dispersion force the particles 
to adopt an almost periodic arrangement. 
All the calculations for different numbers $N$ of particles in the unit cell 
yield very similar results within the statistical uncertainty. 
In particular the elementary transitions illustrated in Fig.~\ref{fig1}  
correspond to relatively large correlation ranges. Moreover, notice
that for small structural disorder $\sigma_R\le5\,\%$ and for small 
$N$ our results for $\overline{\lambda}_\mathrm{c}$ might not be very 
accurate quantitatively, since in these cases the correlation range 
is comparable to the size of the unit cell. 

\begin{figure}
\hspace*{0cm} \includegraphics[width=10cm]{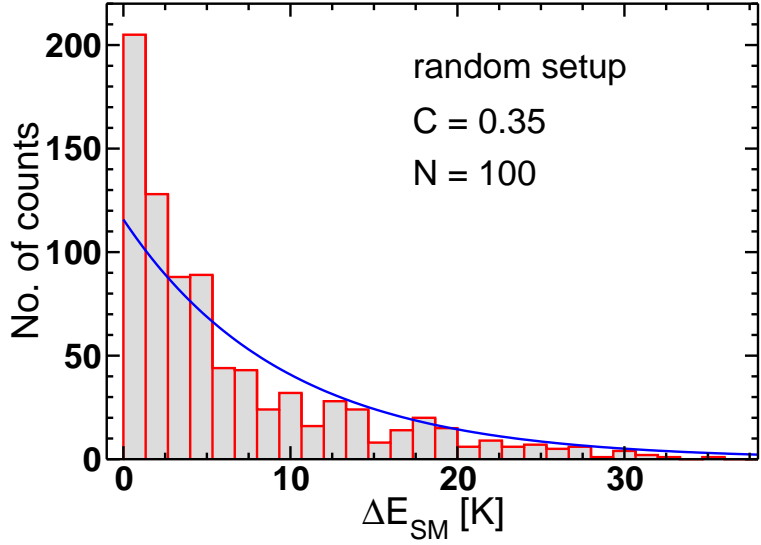} 
\caption{ \label{fig4} 
(Color online) Distribution of the energy barriers $\Delta E_\mathrm{SM}$ 
corresponding to the elementary transitions in a random ensemble having 
a coverage $C=0.35$. The curve shows a fitted exponential law. 
        }
\end{figure} 

The regions where the elementary transitions take place become 
more localized with increasing disorder of the nanostructure. 
Beyond $\lambda_\mathrm{c}$, in the remaining part of the unit cell, 
no significant changes in the directions of the particle magnetizations 
are found. The magnetic transitions 
in strongly interacting particle ensembles can be studied by using 
periodic cells provided that they are significantly larger than 
$\lambda_\mathrm{c}$. For strong disorder already small unit cells 
suffice, whereas for almost periodic ensembles much larger cells are 
required. This renders the simulations far more demanding since 
an important number of particles takes part in an 
elementary transition. In the limit of a periodic arrangement the whole 
system is involved in an elementary relaxation (microvortex state). 
Nevertheless, for moderately disturbed 
ensembles where the underlying square arrangement is still clearly 
recognizable, the correlation range already assumes relatively small 
values [see, for example, Fig.~\ref{fig1}(a)]. It would be therefore
interesting to investigate the dependence of $\lambda_\mathrm{c}$ on 
$\sigma_R$ in more detail, particularly as the periodic lattice is 
approached ($\sigma_R\to 0$). Finally, it should be mentioned that
including single-particle magnetic anisotropies in the calculations
would tend to further reduce $\lambda_\mathrm{c}$, especially for small 
coverages $C$ and large interparticle distances $R_0$. This 
corresponds to weakly interacting particle ensembles, where the 
elementary transitions are essentially 
given by the magnetic reversal of a single particle. 

The energy barriers $\Delta E_\mathrm{SM}=E_\mathrm{S}-E_\mathrm{M}$ 
between a saddle point and the
adjacent minima are one of the central properties governing 
the relaxation dynamics. In Fig.~\ref{fig4} the distribution of 
$\Delta E_\mathrm{SM}$ is shown for a random ensemble having a 
coverage $C=0.35$. The results were
derived taking into account 826 different elementary transitions 
of the same nanostructure. One observes that the probability density
of finding an energy barrier $\Delta E_\mathrm{SM}$ 
follows approximately an exponential behavior, although the precise energy 
dependence of the distribution is not known. 
In any case, it is worth noting that the distribution of barrier energies
contrasts with those of the absolute energies of the saddle 
points and local minima, which were shown to be Gaussian-like for 
random particle ensembles.\cite{JeP02} 

\begin{figure}
\hspace*{0cm} \includegraphics[width=10cm]{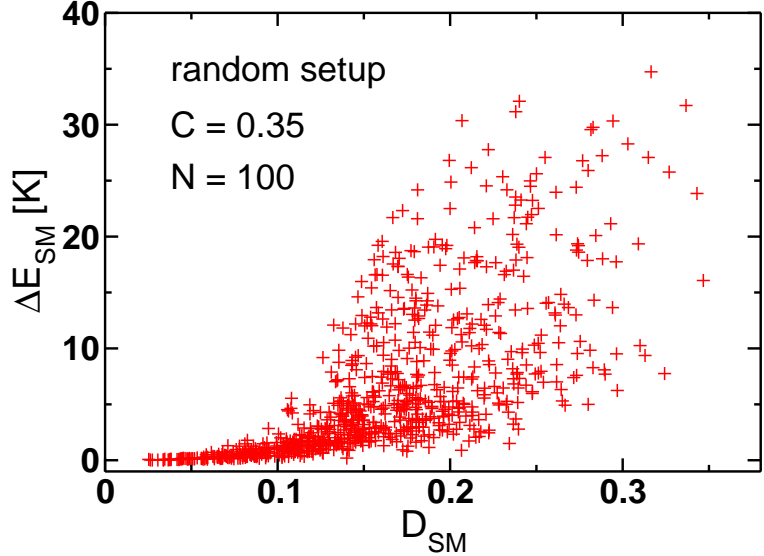} 
\caption{ \label{fig5} 
(Color online) Scatter plot of the energy barrier 
$\Delta E_\mathrm{SM}$ as a function of the distance $D_\mathrm{SM}$ 
between a saddle point and the adjacent local minima. As in 
Fig.~\ref{fig4} the particle setup is random and the coverage is 
$C=0.35$.
        } 
\end{figure} 

Figure~\ref{fig5} shows the correlation between $\Delta E_\mathrm{SM}$ 
and the distance $D_\mathrm{SM}$ between the saddle points and the 
adjacent minima in a random nanoparticle setup [see Eq.(\ref{e2})]. 
As a general trend one observes that if $D_\mathrm{SM}$ is small also 
the corresponding $\Delta E_\mathrm{SM}$ is small. However, for larger 
$D_\mathrm{SM}$ the energy barriers can assume both large and small 
values. Hence, important changes of the magnetic arrangement are
also possible by involving only a single and relatively small 
energy barrier $\Delta E_\mathrm{SM}$. In this connection it should 
be noted that the distribution of the distances 
$D_\mathrm{SM}$ between saddles and minima has been found to follow a 
Poisson-like distribution.\cite{JeP02}

\begin{figure}
\hspace*{0cm} \includegraphics[width=10cm]{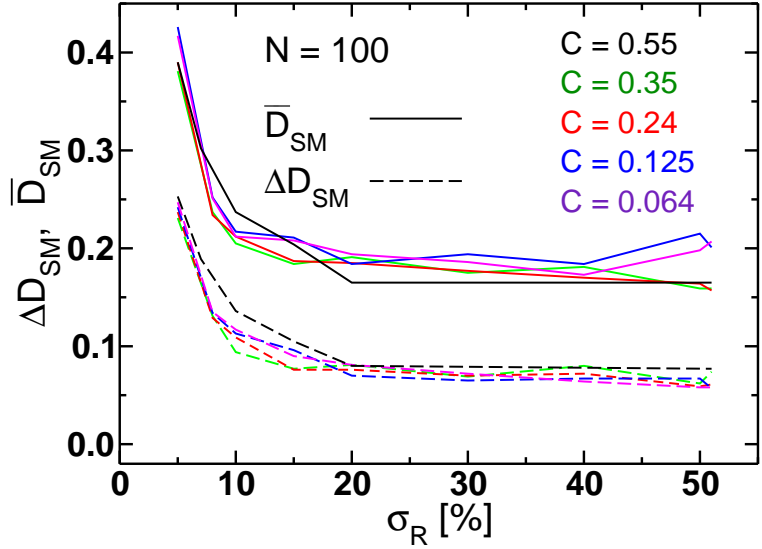} 
\caption{ \label{fig6} 
(Color online) Average distance $\overline{D}_\mathrm{SM}$ 
between saddle points and adjacent minima, as well as the standard 
deviation $\Delta D_\mathrm{SM} = \sqrt{
\langle (D_\mathrm{SM} - \overline{D}_\mathrm{SM})^2\rangle}$ 
as functions of the structural disorder $\sigma_R$.
The results are derived from a large number of elementary transitions
for each different coverage $C$. 
        } 
\end{figure} 

In Fig.~\ref{fig6} the average distance 
$\overline{D}_\mathrm{SM}$ and the standard deviation
$\Delta D_\mathrm{SM} = \sqrt{
\langle (D_\mathrm{SM} - \overline{D}_\mathrm{SM})^2 \rangle}$ 
are given as functions of the structural disorder $\sigma_R$ for 
different coverages $C$. $\overline{D}_\mathrm{SM}$ provides a 
measure of the change in the
magnetic state associated with an elementary transition. 
In agreement with the results for $\lambda_\mathrm{c}$ shown in 
Fig.~\ref{fig3}, $\overline{D}_\mathrm{SM}$ decreases with 
increasing $\sigma_R$, converging to the random-setup limit 
already for $\sigma_R\ge 20\,\%$. No significant dependence on the 
coverage $C$ is found. The distance between the neighboring minima 
of elementary transitions ($D_\mathrm{MM}$) follows closely the 
behavior of $D_\mathrm{SM}$, of course with somewhat larger values. 
As in the case of Fig.~\ref{fig3} the results for weak structural 
disorder $\sigma_R\le5 \%$ are less reliable due to the finite size
of the periodic cell.

\section{Conclusion} 
\label{sec:conc}

The elementary relaxation processes occurring in disordered 2D ensembles 
of dipole-coupled magnetic nanoparticles have been investigated by 
calculating the transitions between neighboring metastable states 
across first-order saddle points. For this purpose an 
eigenvector-following method was implemented.
The extension of the correlated region
where the magnetic directions of the particles vary significantly 
has been calculated. In particular we have 
shown that the magnetic correlation range $\lambda_\mathrm{c}$ becomes 
smaller as the degree of disorder in the nanostructure increases. 
The tendency to localization of elementary transitions in 
disordered magnetic systems renders the simulation of these 
structures computationally less demanding. Moreover, the distribution 
of the energy barriers $\Delta E_\mathrm{SM}$ has been found to follow 
approximately an exponential behavior. An analysis 
of the relation between $\Delta E_\mathrm{SM}$ and the change
$D_\mathrm{SM}$ of the magnetic state shows that 
small energy barriers can lead to both small and large 
magnetic relaxations, whereas large barriers always imply an 
important rearrangement of the magnetic state.

A number of implications and open questions related to our results may
be pointed out. As already noted, the time evolution of the magnetization 
of the nanostructure can be described by a succession of the elementary 
transitions investigated in this work. Particularly important for the 
description of such processes is the knowledge of the connectivity of 
the network of metastable states in the complex spin-glass-like energy 
landscape. For each given system one would have to determine which minima 
are connected by elementary transitions, as well as the corresponding 
transition rates. The latter are mainly defined by the energy barrier,
the eigenfrequencies at the saddle point,  
and the length of the MEP. They can be determined in the framework of 
transition-state theory including finite-temperature effects. 
Besides the elementary transitions involving collective rotations 
of the particles moments within a limited correlation range, the magnetic 
relaxation of the nanostructure as a whole involves some complex 
domain-wall-like motion that results from a sequence of elementary processes. 
In the case of time-dependent fields, averages 
over trajectories should be performed in order to simulate 
hysteresis loops from a microscopic point of view. 

For more realistic simulations and for a detailed comparison with 
experiment it would be interesting to take into account other potentially
important contributions, for example, the magnetic anisotropy energy 
of individual particles and the interaction with external magnetic 
fields. Qualitatively, one expects that increasing the single-particle 
anisotropies or the applied field should tend to reduce the correlation range
of the elementary dynamical process, since the magnetic 
relaxation would be increasingly dominated by single-particle effects. 
Another aspect deserving further study concerns the distribution of 
particle sizes, which is certainly unavoidable in real nanostructures.
In fact previous works\cite{Jen03} suggest that they   
should have similar, probably less severe consequences on the energy 
landscape as the positional disorder considered here. More interesting 
interaction effects on the dynamics are expected to result from other 
types of interparticle couplings, for instance, indirect RKKY 
interactions between NP on metallic substrates, or direct exchange 
interactions at the boundaries of bare NPs in contact. These energy 
contributions can be treated 
straightforwardly using the methods developed in this work.

The trend to increasing localization of the magnetic excitations 
with increasing disorder in the particle arrangement 
should be of interest for tailoring materials for information storage.
In fact, a stronger disorder decreases the minimum distance required for 
an independent switching of neighboring `bits'. Hence, controlling the 
main parameters characterizing the nanostructure allows to optimize 
their magnetic response as potential high-density magnetic-storage 
media or spintronic devices, for example, by tuning the particle 
arrangements and the coverages. This also reveals the limits for 
single-particle read and write processes in interacting magnetic 
nanostructures. Research in these directions is currently in progress. 

\appendix 
\section{Iterative search for saddle points}

\begin{figure}
\hspace*{0cm} \includegraphics[width=10cm]{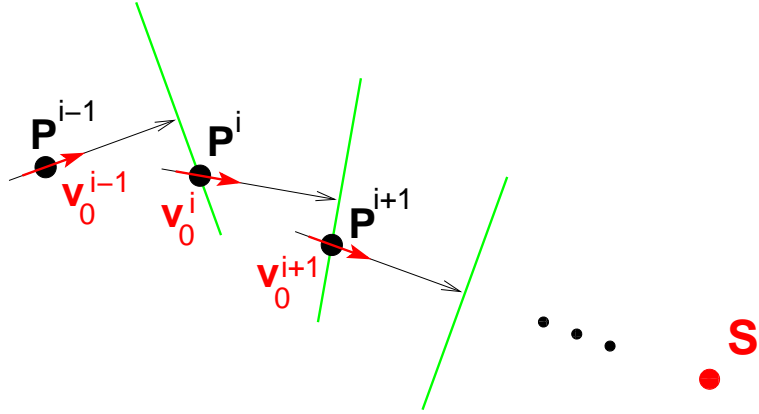}
\caption{ \label{fig7} 
(Color online) Sketch of the eigenvector-following method to determine 
first-order saddle points. 
        } 
\end{figure}

The method to determine the first-order saddle points
is based on the eigenvalues and eigenvectors of the Hessian
matrix of the energy $E$ as a function of the magnetization 
angles $\{\phi_i\}$ [see Eq.~(\ref{e1})].\cite{MoB98} 
The following iteration 
algorithm is applied as sketched in Fig.~\ref{fig7}: 
(i) Consider an arbitrary initial magnetic state $\mathbf{P}^{(k)}$ in the
$N$-dimensional space of the magnetization angles $\{\phi_i\}$. 
(ii) Determine at $\mathbf{P}^{(k)}$ the eigenvector
$\mathbf{v}_0^{(k)}$ of the Hessian corresponding to 
the lowest eigenvalue $\lambda_0^{(k)}$. 
(iii) Maximize the energy of the magnetic state $\mathbf{P}$ 
along $\mathbf{v}_0^{(k)}$ until a state $\mathbf{P}_0$ is found
where the energy shows a maximum along $\mathbf{v}_0^{(k)}$, or 
until a maximal prescribed distance from $\mathbf{P}^{(k)}$ is reached.
(iv) Starting now from $\mathbf{P}_0$ minimize the energy
in the $(N-1)$-dimensional space perpendicular to $\mathbf{v}_0^{(k)}$ 
until the nearest minimum in this restricted subspace is reached. 
(v) Set this minimum as the initial state $\mathbf{P}^{(k+1)}$ of
the next iteration step, and repeat steps (ii)--(iv) until a saddle
point $\mathrm{S}$ is reached with the desired accuracy.
This procedure is similar to the eigenvector-following method described 
in Ref.~\onlinecite{Wales00}. 

The convergence criterion at a first-order saddle point $\mathbf{S}$ 
requires that the norm of the gradient be sufficiently small, 
that the lowest eigenvalue $\lambda_0$ at $\mathrm{S}$ be negative, and 
that all other eigenvalues be positive. The two corresponding minima
$\mathrm{M}_1$ and $\mathrm{M}_2$ adjacent to a saddle point are then
determined by descending from the saddle point along the two opposite 
directions defined by the eigenvector $\mathbf{v}_0$ corresponding 
to $\lambda_0$. The resulting MEP is obtained by following
the gradient by steepest descent.

\end{document}